\documentclass[sigconf]{acmart}
\AtBeginDocument{%
  }

\usepackage[utf8]{inputenc}
\usepackage[linesnumbered,ruled,lined]{algorithm2e}
\usepackage{subfigure}
\usepackage{multirow}
\usepackage{threeparttable}
\usepackage{graphicx}

\usepackage{amsmath}  
\newtheorem{definition}{Definition}

\long\def\comment#1{}

\setcopyright{acmlicensed}
\copyrightyear{2026}
\acmYear{2026}
\acmDOI{XXXXXXX.XXXXXXX}
\acmConference[DAC'26]{the 63st ACM/IEEE Design Automation Conference}{July 26--29,
  2026}{Long Beach, CA, USA}
  
\acmISBN{978-1-4503-XXXX-X/2018/06}

\begin{document}

\title{LHGstore: An In-Memory Learned Graph Storage for Fast Updates and Analytics}


\author{Pengpeng Qiao}
\affiliation{%
  \institution{Institute for Science Tokyo}
  \city{Tokyo}
  \country{Japan}
  }
\email{peng2qiao@gmail.com}

\author{Zhiwei Zhang}
\affiliation{%
  \institution{Beijing Institute of Technology}
  \city{Beijing}
  \country{China}
  }
\email{zwzhang@bit.edu.cn}

\author{Xinzhou Wang}
\affiliation{%
  \institution{Beijing Institute of Technology}
  \city{Beijing}
  \country{China}
  }
\email{xzwang@gmail.com}

\author{Zhetao Li}
\affiliation{%
  \institution{Jinan University}
  \city{Guangzhou}
  \country{China}
  }
\email{liztchina@gmail.com}

\author{Xiaochun Cao}
\affiliation{%
  \institution{Sun Yat-sen University}
  \city{Shenzhen}
  \country{China}
  }
\email{caoxiaochun@mail.sysu.edu.cn}

\author{Yang Cao}
\affiliation{%
  \institution{Institute for Science Tokyo}
  \city{Tokyo}
  \country{Japan}
  }
\email{cao@c.titech.ac.jp}


\renewcommand{\shortauthors}{Qiao et al.}

\begin{abstract}
Various real-world applications rely on in-memory dynamic graphs that must efficiently handle frequent updates while supporting low-latency analytics on evolving structures. Achieving both objectives remains challenging due to the trade-off between update efficiency and traversal locality, particularly under highly skewed degree distributions. This motivates the design of graph indexing schemes optimized for in-memory graph management on modern multi-core CPUs. We present \textbf{LHGstore}, a degree-aware \textbf{L}earned \textbf{H}ierarchical \textbf{G}raph storage that, for the first time, integrates learned indexing into graph management. LHGstore designs a two-level hierarchy that decouples vertex and edge access and further organizes each vertex’s edges using data structures adaptive to its degree. Lightweight arrays are used for low-degree vertices to maximize traversal locality, while learned indexes are applied to high-degree vertices to improve update throughput. Extensive experiments show that LHGstore achieves 5.9-28.2× higher throughput and significantly faster analytics than SOTA in-memory graph storage systems.
\end{abstract}

\maketitle

\section{Introduction}
Many real-world graphs~\cite{baranwal2023optimality,chang2022efficient} are widely used in applications, such as social networks, recommendation systems, and fraud detection~\cite{khan2024data}. 
These graphs evolve dynamically over time, while applications simultaneously require graph analytics to extract insights~\cite{yang2023fast,sallinen2023real,mo2024vldb}. 
For example, in a financial network, where accounts are nodes and transactions are edges, the graph updates dynamically with new or canceled transactions, and algorithms such as Breadth-First Search (BFS) are applied to trace suspicious transaction paths.
Therefore, designing an efficient graph storage structure to support both requirements (\underline{\textbf{R$_1$}}) high-throughput graph updates (\emph{e.g.,} edge insertions and deletions), and (\underline{\textbf{R$_2$}}) high-performance graph analytics (\emph{e.g.,} BFS, SSSP) is a significant problem. That is, both \emph{graph updates} time and \emph{graph analytics} time must be fast.

\begin{figure}[tp]
    \centering
    \includegraphics[scale=0.3]{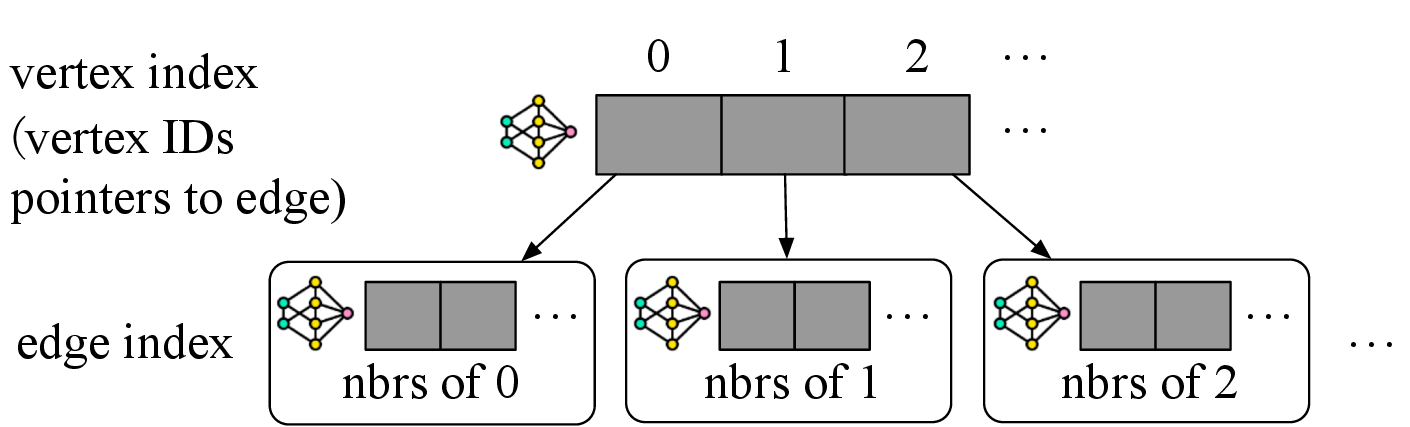}
    \vspace{-0.2cm}
    \caption{ A high-level architecture for graph indexing. A vertex index keeps track of where the neighbors (nbrs) for all vertices, and a list of edge indexes for each vertex’s edges.}
    \label{fig:intro}
    \vspace{-0.5cm}
\end{figure}

Several existing in-memory dynamic graph storage systems~\cite{chu2024livak,LiveGraph,Teseo,Terrace,Sortledton,shi2024spruce,mo2025aster} struggle to satisfy both requirements simultaneously. 
\emph{First}, traditional indexing structures suffer from inherent inefficiencies. For example, B+trees~\cite{gao2024revisiting} or ART~\cite{art} rely on costly key comparisons, making both suboptimal for graph updates. 
\emph{Second}, some systems prioritize update efficiency but compromise traversal locality, resulting in poor analytics performance~\cite{dhulipala2019low,zhang2023reachability}. 
For example, CSR-like layouts~\cite{shun2013ligra} achieve excellent sequential locality for analytics but make dynamic updates prohibitively expensive.
More recent efforts, such as Teseo~\cite{Teseo}, LiveGraph~\cite{zhu2020livegraph}, and LSMGraph~\cite{yu2024lsmgraph} improve specific aspects through packed memory arrays, log-based storage, or LSM-tree, yet they still rely on \emph{comparison-heavy} or \emph{shift-heavy} data structures whose update or lookup costs remain substantial. This gap highlights the need for a new indexing that can optimize both updates and analytics.

\textbf{Trade-off.}
Designing graph indexing for dynamic graphs faces a fundamental trade-off between update efficiency and traversal locality. High-throughput updates (\underline{\textbf{R$_1$}}) require efficient \emph{lookup} and \emph{insertion} operations with minimal data movement, favoring discontiguous layouts.
In contrast, high-performance analytics (\underline{\textbf{R$_2$}}) such as BFS rely on contiguous memory layouts that maximize CPU cache locality. These two requirements are inherently conflicting: discontiguous layouts reduce update cost but fragment memory access, while contiguous layouts favor analytics but require expensive element shifts during edge insertions and deletions. \textbf{Therefore, it is non-trivial for a traditional graph indexing to support both efficient updates and analytics in dynamic graphs.}

Recent advances in learned indexes~\cite{18sigmod-rmi,20sigmod-alex,sun2023learned,21vldb-lipp,chen2024oasis,qiao2024astore} show that lightweight machine learning models can replace \emph{comparison-heavy} indexes with prediction-driven data placement. 
By leveraging key distributions, learned indexes achieve near-$O(1)$ lookup and insertion performance while preserving ordered layouts, exactly the properties desirable for in-memory systems on modern multi-core CPUs, where cache locality dominates performance.
Moreover, graph adjacency lists can also be naturally modeled as key–value pairs, where the key is the vertex ID and the value is the neighbor ID, making learned indexes a promising candidate for dynamic graph storage. 
However, directly applying learned indexes to graphs faces inherent limitations. For example, when many edges of a high-degree vertex are predicted to the same position, lookups require additional scans and insertions may trigger large shifts, leading to $O(deg(u))$ costs. This overhead prevents learned index from fully realizing their benefits in dynamic graph.

To address the above limitations, we propose {\bf LHGstore}, a degree-aware {\bf L}earned {\bf H}ierarchical {\bf G}raph indexing framework. 
As illustrated in Fig.~\ref{fig:intro}, LHGstore adopts a two-level hierarchical architecture that decouples vertex lookup from neighbor access.
This allows adjacency lists to benefit from near-constant-time predictions while preserving ordered layouts that are essential for graph analytics.
Specifically, we model graphs as a set of key-value pairs and exploit learned index~\cite{20sigmod-alex} as the core submodule supporting our hierarchical graph storage.
LHGstore introduces a secondary edge-level index with a learned index for each vertex, which theoretically reduces both lookup and insertion costs to expected $O(1)$, thereby satisfying requirement (\underline{\textbf{R$_1$}}).
Furthermore, real-world dynamic graphs often follow skewed vertex degree distributions, with a few high-degree vertices and many low-degree vertices, as shown in Table~\ref{tab:degree}.
Using a uniform “one-size-fits-all” index design for each vertex is therefore suboptimal.
The ideal structure for storing a vertex’s neighbors depends on the access pattern of graph analytics and the cost of doing graph updates. This motivates a degree-aware design. 
For low-degree vertices, we use unsorted arrays, which avoid unnecessary model overhead and preserve sequential locality for analytics (\underline{\textbf{R$_2$}}). For high-degree vertices, we employ edge-level learned indexes, which maintain efficient updates and traversals.
In summary, the contributions of this paper are as follows.

\begin{table}
\begin{center}
\newcommand{\tabincell}[2]{\begin{tabular}{@{}#1@{}}#2\end{tabular}}
\caption{Distribution of degree of vertices. }
\vspace{-0.3cm}
\resizebox{1\linewidth}{!}{
\begin{tabular}{cccccc}
      \hline
      \emph{Graph} &$\leq10$\emph{nbrs}  &$\leq100$\emph{nbrs}  &$\leq1000$\emph{nbrs} & {\tabincell{c}{Avg. \\ degree}} & {\tabincell{c}{Max \\ degree}}\\
      \hline
      Twitter & 64.5\% & 95.4\% & 99.51\% & 39 &2997469\\
      LiveJournal & 65\% & 97.2\% & 99.98\%  & 17 &14815\\
      \hline
    \end{tabular}}
\label{tab:degree}
\end{center}
\vspace{-0.55cm}
\end{table}

\noindent \underline{\emph{New Perspective}.} 
This paper is the first attempt to introduce learned index as a solution for considering \emph{update efficiency} and \emph{traversal locality} to explore in-memory dynamic graph storage.

\noindent \underline{\emph{Simple yet Effective Solution.}} 
We present LHGstore, a degree-aware learned hierarchical graph indexing structure that decouples vertex and neighbor access while assigning different adjacency layouts based on vertex degree.
This design exploits skewed degree distributions in real-world graphs to minimize data movement for updates and maximize locality for analytics.

\noindent \underline{\emph{Extensive experiments}.} 
We conduct experiments with three workflows and five graph analytics algorithms on synthetic and real-world graphs, demonstrating that our approaches process graph updates and analytics performance efficiently.

\section{Related Works}
\label{sec:related}
\textbf{Graph Storage.} 
A classic graphs storage structure is CSR \cite{csr,yang2024leveraging}, which provides high analytics performance for static graphs by storing both vertices and edges continuously. 
LLAMA \cite{llama} extends CSR-like structures to dynamic settings using multi-versioned snapshots. Teseo \cite{Teseo} employs PMA-based storage with transactional support, Aspen \cite{dhulipala2019low} leverages copy-on-write PAM trees to optimize reads, LiveGraph \cite{zhu2020livegraph} adopts TEL blocks for efficient insertions, and Sortledton \cite{Sortledton} integrates hash tables with skip lists to balance update and query trade-offs.   
However, existing approaches primarily rely on traditional index structures for dynamic graph data storage, which often face significant performance bottlenecks.

\noindent\textbf{Learned Storage.} 
ML-based storage~\cite{18sigmod-rmi} has emerged as a promising research direction, and several theoretical works have been proposed in this area~\cite{yang2024lits,kraska2019sagedb,luo2018optimizing,ferragina2020learned,marcus2020cdfshop,li2021finedex,yu2024camal,li2024gaussml,hu2022selectivity} using ML techniques.
\cite{18sigmod-rmi} proposes the recursive model index, which uses a few levels of models to partition the dataset.
Ding et al. \cite{20sigmod-alex} propose ALEX, the first learned index that can support dynamic insertion and introduces a top-down adaptive way of constructing a tree-shaped linear learned index.
However, there is no work for graph.

\section{Motivation and Baseline}
\label{section:lgstore}

\subsection{Motivation}
\subsubsection{Memory Access Patterns}
Graph analytics algorithms are known to be memory access bound~\cite{Sortledton}. Most graph algorithms, including those benchmarked in Graphalytics~\cite{graphalytics-benchmark}, follow either a sequential or random access pattern. 
Algorithm~\ref{algo:iterupdate} shows the sequential pattern: the algorithm iterates over all vertices $V$, traverses their adjacency lists, and updates vertex states based on their neighbors. This pattern is typical of iterative algorithms such as PageRank.  
By contrast, random access patterns are determined at runtime. 
\vspace{-0.25cm}
\begin{algorithm}[htp]
\caption{\emph{Sequential access pattern}} \label{algo:iterupdate}
\SetAlgoVlined
\While{condition}{
  \ForEach{$v \in V$}{
    \ForEach{$u \in v.nbrs$}{
      $state[v] \leftarrow f_{\emph{update}}(state[v],\, v,\, u)$\;
    }
  }
}
\end{algorithm}
\vspace{-0.4cm}

In main-memory graph engines, the performance of this sequential pattern is almost entirely determined by the underlying storage layout. If the edges of a vertex are not placed contiguously, traversal locality is lost, causing frequent cache misses. This issue is particularly severe for high-degree vertices, where fragmented memory layouts amplify random DRAM accesses.
Optimizing storage to better match access patterns is essential for accelerating analytics.

\begin{figure}[htp]
\vspace{-0.25cm}
    \includegraphics[scale=0.6]{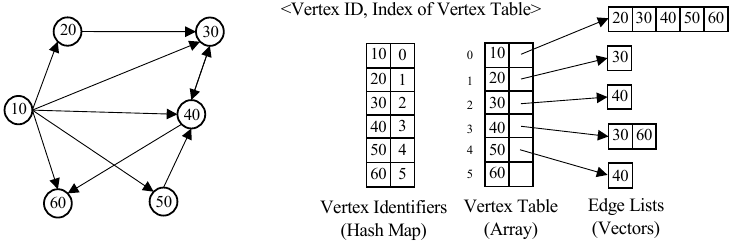}
    \vspace{-0.3cm}
    \caption{A typical graph storage.}
\label{fig:graph_store}
\vspace{-0.45cm}
\end{figure}

\subsubsection{Typical Graph Storage} 
Typical in-memory graph storage~\cite{zhu2020livegraph} commonly adopt a vertex table-based architectureconsisting of a hash map, a static vertex table, and adjacency lists (Fig.~\ref{fig:graph_store}). 
While this structure enables $O(1)$ vertex lookups, it faces two key limitations for dynamic graphs
First, the static vertex table restricts scalability and complicates dynamic vertex insertions or deletions. Second, adjacency lists lack internal indexing, so edge \emph{lookups} and \emph{inserts} degrade to costly linear scans. 
Together, these issues make such designs increasingly insufficient for high-performance in-memory dynamic graph workloads, where random pointer chasing and scattered adjacency lists are prohibitively expensive.

\noindent\textbf{Bridging the trade-off.} 
Building on the motivation above, a central problem in dynamic graph storage lies in balancing two competing requirements: (\underline{\textbf{R$_1$}}) efficient updates and (\underline{\textbf{R$_2$}}) locality-preserving traversals.
Contiguous layouts minimize access overhead for analytics but incur expensive data shifts on updates. Discontiguous layouts reduce update costs but destroy locality, leading to poor traversal performance. Designing an indexing that supports both fast updates and analytics on dynamic graphs is thus challenging.  

\noindent\textbf{Our focus.}
In this paper, we aim to tackle this tension head‐on by exploring learned index and hybrid data layout strategies.

\subsection{Our Baseline Design}

\begin{definition}
\label{def:graph}
\noindent{\bf \emph{Graph to key-value pairs}.} A graph $G=(V, E)$ is a set of vertices $V$, a set of edges $E$. We denote the number of vertices with $|V|$, the number of edges with $|E|$, and the degree of a vertex $v \in V$ with $deg(v)$. Each edge is a key-value pair, $(u, v)$ where $u, v \in V$. For instance, for each edge $(u, v)$, $u$ is the key and $v$ is the value. Each vertex $v \in V$ is represented by a unique non-negative integer less than $|V|$ (i.e. $v \in \{1, 2, ..., |V|\}$).
\end{definition}

\begin{figure}[!t]
    \includegraphics[scale=0.22]{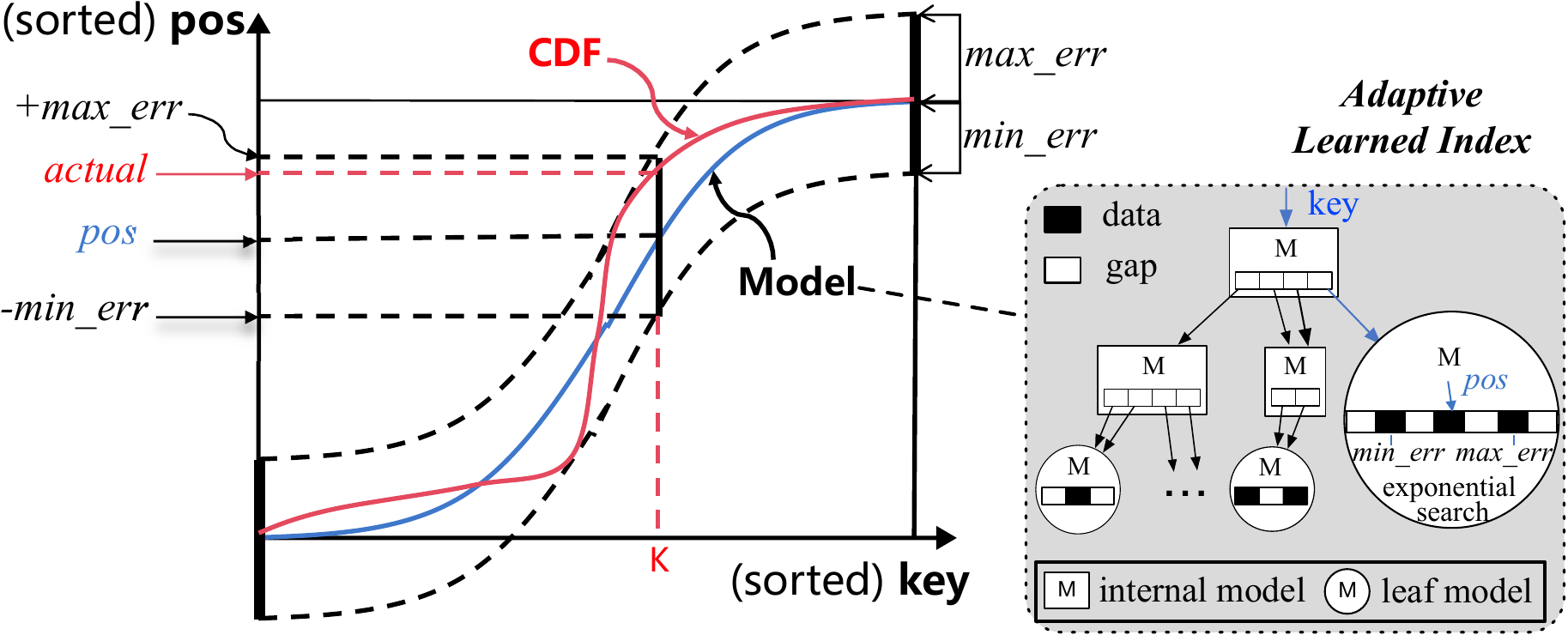}
    \vspace{-0.4cm}
    \caption{An example of using ML models to predict the position within a sorted array for a given key.}
    \label{fig:alex}
    \vspace{-0.4cm}
\end{figure}

\noindent{\bf Our baseline design: Learned graph storage (LGstore).}
The key idea behind our LGstore is to leverage ML models~\cite{20sigmod-alex} to learn the distribution of graphs. 
Based on the Definition~\ref{def:graph}, graph data can be seen as a set of key-value pairs. 
For instance, for each edge $(u, v)$, $u$ is the key and $v$ is the value. 
The structure of LGStore is derived from the adaptive learned index~\cite{20sigmod-alex}, as shown in Fig.~\ref{fig:alex}.
LGStore is a recursive model index (RMI~\cite{itai1981sparse}), in which all the models are organized in a tree structure.
Each non-leaf node, referred to as \emph{internal node}, contains a learned model and pointers to its sub-nodes. 
Whenever it needs to identify the position of a given key (vertex), the internal node uses its model to determine the sub-node that needs to be searched. 
In fact, each internal node manages a certain range of vertices, and the model in it divides all vertices into partitions as many as the number of its sub-nodes.
Each \emph{leaf node} plays the role of managing edges as well as keeping a learned model, which maps a vertex to a position.
The leaf node uses gapped arrays~\cite{bender2006insertion} to store the edges of each vertex.

\begin{figure}[!t]
    \includegraphics[scale=0.45]{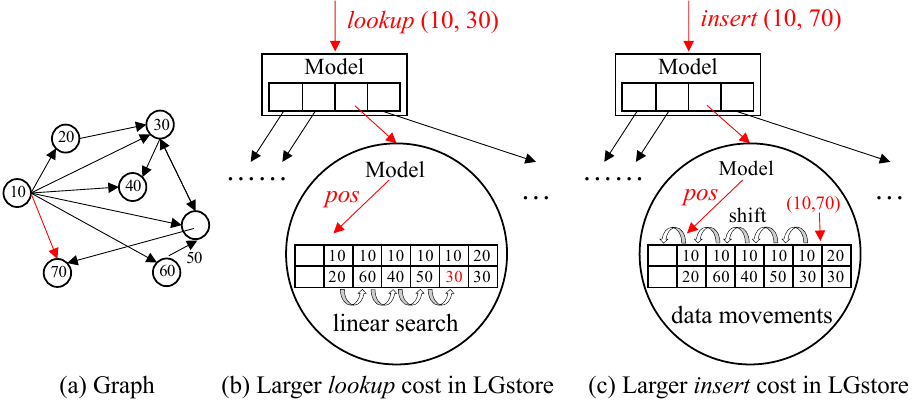}
    \vspace{-0.3cm}
    \caption{An example of a leaf node in LGstore.}
    \label{fig:kv_storage}
    \vspace{-0.5cm}
\end{figure}

\begin{example}
\label{example:lgstore}
\noindent{\bf \emph{LGstore stores graph}.} We use LGStore to store a dynamic graph, as shown in Fig. \ref{fig:kv_storage}. All edges of the vertex 10 are continuously stored in a leaf node in an unsorted way, as shown in Fig. \ref{fig:kv_storage}(b). When we lookup the edge $(10, 30)$, the linear model of the leaf node will predict the position of $(10, 20)$, then we have to do a linear search over 5 elements. As shown in Fig. \ref{fig:kv_storage}(c), when we insert a new edge $(10, 70)$, we need to shift 5 elements to place the new edge in the right position. 
\end{example}

\section{LHGstore Design}
\vspace{-0.2cm}
\label{sec:storage}
\subsection{Limitations of LGstore}
Although LGstore represents a natural first step in applying learned index to graphs, it suffers from two limitations.

\noindent\textbf{\emph{Limitation-1}: $O(deg(u))$ costs for lookups and inserts.}
As illustrated in Example~\ref{example:lgstore}, when multiple edges of the same vertex are predicted to the same position, they are stored contiguously without gaps. In this case, looking up an edge requires a linear scan over all adjacent edges of that vertex, and inserting a new edge requires shifting multiple elements. The theoretical worst-case time complexity of both \emph{lookup} and \emph{insert} for a vertex $u$ is $O(deg(u))$.
For high-degree vertices in an in-memory setting, such O(deg(u)) behavior translates into long cache line walks and DRAM bursts, severely limiting throughput.

\noindent\textbf{\emph{Limitation-2}: One-size-fits-all layout under skewed degree distribution.}  
Table~\ref{tab:degree} presents the distribution of vertex degrees in real-world graphs that exhibit skewness. 
Considering skewed graph, LGstore takes “one-size-fits-all” data structure design, which leaves performance on the table when processing and updating skewed graphs.
If a vertex has a low degree, a simple data structure incurs minimal indirection and supports efficient traversal and updates. If a vertex has a high degree, however, a more complex data structure with better asymptotic search and update performance may be more suitable. Therefore, it is necessary to treat low- and high-degree vertices differently to achieve better performance.

\subsection{Overview of LHGstore}
\begin{figure}[!t]
    \centering
    \hspace{-0.2cm}
    \includegraphics[scale=0.5]{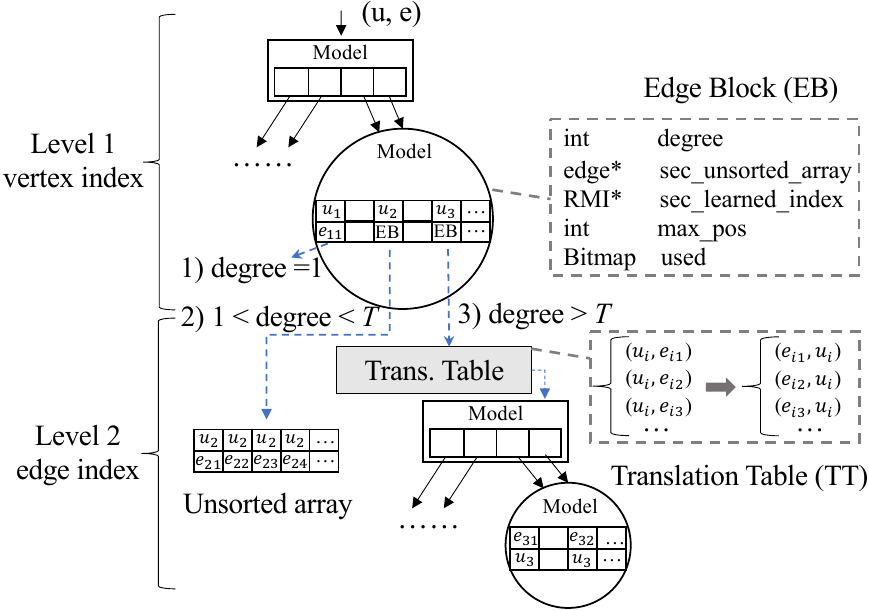}
    \vspace{-0.3cm}
    \caption{The structure of LHGstore.\label{fig:arch}}
    \vspace{-0.5cm}
\end{figure}

We propose LHGstore to address the above issues in LGstore.
The architecture of LHGstore is illustrated in Fig.~\ref{fig:arch}.

\noindent\textbf{Hierarchical design.}  
To overcome \emph{Limitation-1}, LHGstore adopts a two-level hierarchical index to eliminate the impact of the same position from model predictions.
At the first-level (vertex index), LHGstore uses a learned index to store the edges of each vertex, but only if its degree$=$1. Otherwise, the value of each vertex (degree$>$1) is an edge block that points to the corresponding second-level index (edge index).
At the second-level, each edge index adopts another learned index to store all edges of a vertex, ensuring ordered adjacency for locality and efficient updates through gapped array.
This design reduces both lookup and insertion costs for a vertex $u$ from $O(deg(u))$ in LGstore to expected $O(1)$ in LHGstore.

\noindent{{\bf Degree-aware design.}}
To address \emph{Limitation-2}, LHGstore employs a hybrid strategy in which the layout of a vertex should depend on that vertex’s degree.
For the different degrees of vertex, array-based and learned-based index structures provide different guarantees and exhibit crossover points in terms of updateability and locality. 
A degree threshold $T$ is used to distinguish between the two layouts (see Section~\ref{subsec:edge table} for analysis of the \emph{crossover point}).

\subsection{Vertex Index}
\label{subsec:vertex table}

As shown in Fig.~\ref{fig:arch}, we use the adaptive learned index introduced in LGstore for vertex-level index.  

\textbf{\emph{Edge block.}} The edge block corresponding to vertex $u_i$ stores the degree of $u_i$, two pointers that point to the edge index, named \emph{sec\_unsorted\_array} and \emph{sec\_learned\_index}, and two auxiliary variables for the edge index, named \emph{max\_pos} and \emph{used}.
Edge blocks are ordered by vertex index.

Let $\{e_{i1}, e_{i2}, ... \}$ denote neighbor vertices of $u_i$.
When the $deg(u_i)$ is equal to 1 (e.g., $u_1$), the first level directly stores its neighbor (e.g., $e_{11}$) in-place without the need for a second-level index.
When the $deg(u_i)$ is greater than 1 (e.g., $u_2, u_3$), we use \emph{edge block} containing necessary metadata for $u_i$.

 For vertex index, the key is the vertex ID $u_i$, and the value is a pointer to the corresponding adjacency block (edge index) or its edge $e_i$.
When an edge of vertex $u_i$ is inserted or deleted, we first update the degree of vertex $u_i$ and then perform the corresponding operation in the second-level index.

\subsection{Edge Index}
\label{subsec:edge table}

\begin{figure}
\vspace{-0.2cm}
 \centering
 \vspace*{-0.15cm}
 \begin{tabular}[t]{l}
    \hspace{-0.5cm}
  \subfigure[lookup]{
   \raisebox{0.03\height}{\includegraphics[width=0.43\linewidth]{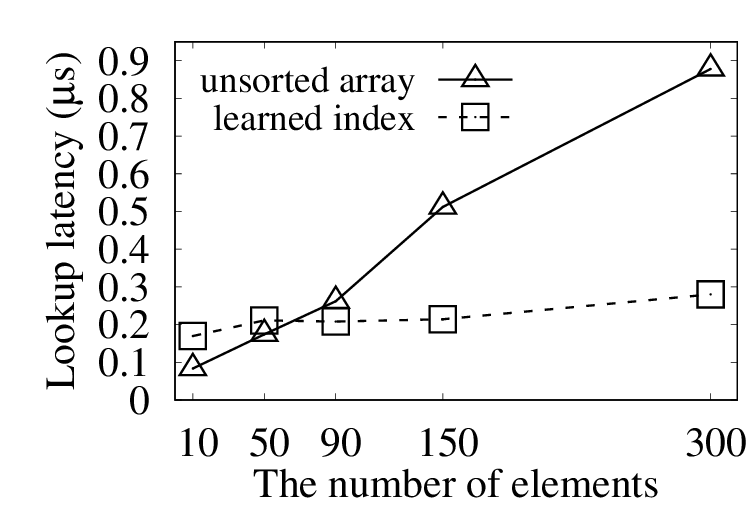}}
  }
      \hspace{0.4cm}
  \subfigure[insert]{
   \raisebox{0.03\height}{\includegraphics[width=0.43\linewidth]{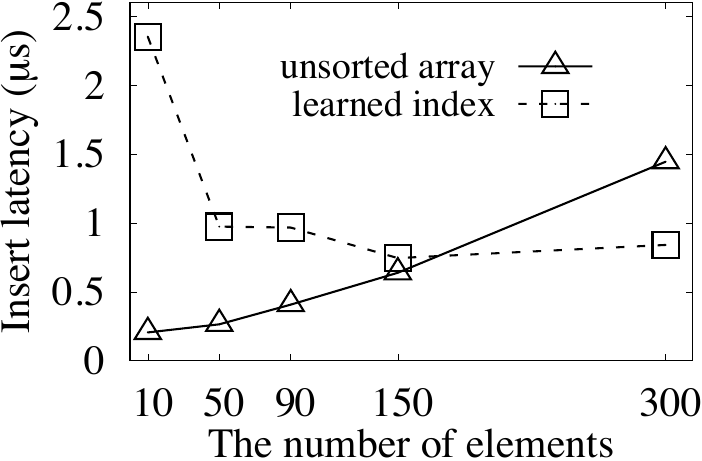}}
  }
 \end{tabular}
  \vspace*{-0.4cm}
 \caption{The latency of \emph{lookup} and \emph{insert} operations on unsorted array and learned index.}
  \vspace*{-0.3cm}
 \label{fig:micro benchmark}
\end{figure}

\noindent{{\bf Balancing locality and updateability.}}  
The array-based and learned index-based data structures provide different guarantees in terms of locality and updatability.
Learned indexes are quick to update, which offers good performance for graph updates. 
In contrast, an array-based structure has worse updatability than learned index, but supports a fast graph analytics algorithm because edges are stored contiguously in memory with better data locality. 
In fact, there is a crossover point in the update performance of learned index and array depending on the degree of the vertex.
To validate this, we conducted a micro-benchmark comparing lookup and insertion latency in arrays and learned indexes. 
The results, shown in Fig.~\ref{fig:micro benchmark}, reveal a clear \emph{crossover point}: for $n<70$, the lookup in the unsorted array is faster than the
learned index, and for $n<150$, the insertion in the unsorted
array surpasses that in the learned index.
This identifies a crossover point that serves as a principled guideline for selecting the index threshold $T$ that determines whether the edges of each vertex are stored in unsorted array or learned index.

For each high-degree vertex ($d(u)>T$), LHGstore assigns a learned index for each vertex as the edge index to ensure efficient lookup and insertion. 
If we directly use the vertex ID $u_i$ as the key for all edges of $u_i$, then all key-value pairs have the same key, and the model would predict the same position, making the learned index ineffective. To solve this, we introduce a translation table so that the neighbor ID $e_i$ is treated as the key and $u_i$ is the value.
For instance, we first transform all edges $\{(u_i, e_{i1}), (u_i, e_{i2}),...\}$ into $\{(e_{i1}, u_i), (e_{i2}, u_i),...\}$ by translation table, and then build the learned index. For each low-degree vertex ($d(u) \leq T$), unsorted array is adopted instead, which supports constant-time updates and preserves sequential scans for analytics. All learned index modules are implemented using adaptive linear regression models~\cite{20sigmod-alex}, which can be trained efficiently with negligible overhead.

\begin{example}
\label{example:lhgstore}
\noindent{\bf \emph{LHGstore stores graph}.} Fig.~\ref{fig:arch} illustrates how LHGstore stores vertices of different degrees in a dynamic graph. Consider the graph shown in Fig. 5(a). We assume that the threshold is $T=5$. Vertex $v_{20}$ has degree 1, so its single neighbor is stored directly in the vertex index without an edge-level structure. Vertex $v_{30}$ has degree 2 ($1<deg(v_{30})<T$), and its neighbors are placed in an unsorted array that supports efficient insertions by placing new neighbors into free slots. Vertex 10 has 5 edges $\{v_{20}, v_{30}, v_{40}, v_{50}, v_{60}\}$. Because $deg(v_{10}) > T$, its edge block is managed by an edge-level learned index. For example, when inserting a new edge $(v_{10}, v_{70})$, the model predicts its approximate position after 70 and a small correction finalizes the insertion,  reducing insertion complexity to expected near-$O(1)$.
\end{example}

\begin{algorithm}[htp]
\caption{\emph{findEdge}$(u,v)$}     \label{algo:find}
\emph{pos}$(u)$ $\leftarrow$ VIndex.predict($u,v$) ; \\
\SetAlgoVlined
\If{\emph{pos}$(u).value == null$}{ 
    {\bf Return} False  \ \ \ \  \ \ \ \ /$*$ not exist vertex $u$ $*$/ \\
}
\ElseIf{\emph{pos}$(u).value == v$}{
    {\bf Return} True  \ \ \ \  \ \ \ \ /$*$ $deg(u) = 1$ $*$/ \\    
}
\Else{
    \emph{EB}($u$) = \emph{pos}($u$).value ; \ \ \ \  \ \ \ \ /$*$ $deg(u) > 1$ $*$/ \\
    \If{\emph{EB}$(u).degree > T$}{ 
        EIndex = \emph{EB}$(u)$.$sec\_learned\_index$ ;  \\
        ($v,u$) = {\bf Translation\_Table}($u,v$); \\
        $value \leftarrow$ EIndex.predict($v,u$).value ; \\
    }
    \Else{
        $value \leftarrow$  \emph{EB}$(u)$.$unsorted\_array$.{\bf Traversal} ; \\
    }
    \If{$value == null$}{
        {\bf Return} False  \ \ \  \ \ \ \ /$*$ not exist the edge ($u$, $v$) $*$/ \\
    }
    \Else{
        {\bf Return} True  \\
    }
}
\end{algorithm}

\subsection{Operations}
\label{sec:op}

{\bf  \emph{findEdge$(u,v)$}.} The procedure of \emph{findEdge$(u,v)$} is shown in Algorithm~\ref{algo:find}.
findEedge$(u,v)$ returns whether edge $(u,v)$ exists in the graph. 
Given an edge $(u,v)$, LHGstore first uses vertex index (VIndex) to predict the position where $u$ belongs, and then gets the payload of $u$, denoted as \emph{pos}$(u)$ (line 1).
LHGstore checks the existence of $u$ (line 2) or whether the edge $(u, v)$ is stored in VIndex (line 4). 
If the adjacent edges of $u$ are stored in the edge block (\emph{EB}), we use \emph{sec\_learned\_index} to predict the position of $v$ when degree $> T$ (lines 7-11) . Otherwise, we traverse the unsorted array in the range of $0\sim$ \emph{EB}$(u).max\_pos$ to check whether $v$ exists (lines 12-17).

\begin{algorithm}[htp]
\caption{\emph{insertEdge}$(u,v)$}     \label{algo:insert}
\emph{pos}$(u)$ $\leftarrow$ VIndex.predict($u,v$) ; \\
\SetAlgoVlined
\If{\emph{pos}$(u).value == null$}{ 
    insert $(u, v)$ into VIndex ;  \ \ \ \ /$*$ not exist vertex $u$ $*$/ \\
}
\ElseIf{\emph{pos}$(u)$.value is not \emph{EB}}{
    construct \emph{EB}$(u)$; \\
    \emph{EB}$(u)$.degree = 2 ; \\
    \If{$T == 2$}{ 
        \emph{EB}$(u).sec\_learned\_index$.{\bf insert}(all edges) ; \\
    }
    \ElseIf{$T > 2$}{
        \emph{EB}$(u).sec\_unsorted\_array$.{\bf insert}(all edges) ; \\
        update \emph{EB}$(u).used$, \emph{EB}$(u).max\_pos$ ; \\
    }
    update $(u$,\emph{EB}$\left(u\right))$ into VIndex ; \\
}
\Else{
    \emph{EB}$(u).degree + 1$ ; \\
    \If{\emph{EB}$(u).degree \geq T$}{ 
        \emph{EB}$(u).sec\_learned\_index$.{\bf insert}($u,v$) ; \\
    }
    \ElseIf{\emph{EB}$(u).degree < T$}{
            \emph{EB}$(u).sec\_unsorted\_array$.{\bf insert}($u,v$) ; \\
            update \emph{EB}$(u).used$, \emph{EB}$(u).max\_pos$ ; \\
    }
}
\end{algorithm}

{\bf  \emph{insertEdge$(u,v)$}.} In Algorithm~\ref{algo:insert},  \emph{insertEdge$(u,v)$} inserts an edge $(u,v)$ into the graph. Similar to $findEdege(u,v)$, LHGstore first get \emph{pos}$(u)$ from the VIndex (line 1). 
If $u$ does not exist, it is directly inserted as a new edge $(u,v)$ into VIndex (line 3).
If \emph{EB}$(u)$ does not exist (i.e., the degree of $u$ is equal to 1), LHGstore needs to construct \emph{EB}$(u)$ with $deg(u)$=2, where the decision of whether to build the edge index of $u$ as $sec\_learned\_index$ or $sec\_unsorted\_array$ based on the size of $T$ (lines 4-11). For example, if $T=2$, the \emph{EB}$(u)$ is constructed using a learned index (line 8), otherwise it is constructed using an unsorted array (lines 10-11). And then LHGstore updates the edge block \emph{EB}$(u)$ into the VIndex (line 12).
If \emph{EB}$(u)$ exists, there are two cases when inserting $(u, v)$ into the existing edge block.
(1) When \emph{EB}$(u)$.degree is greater than or equal to $T$, LHGstore uses $sec\_learned\_index$ to store the edge $(u, v)$ (line 16).
(2) The degree of $u$ is less than $T$, then LHGstore inserts $(u, v)$ into the first empty position of the unsorted array, and updates $EB(u).used$ and $EB(u).max\_pos$ (lines 18-19).

{\bf  \emph{deleteEdge$(u,v)$}.}
The process of \emph{deleteEdge$(u,v)$} is similar to the insertion operation. LHGstore first gets the position of $(u,v)$ and then deletes the edge $(u,v)$ from VIndex or $sec\_learned\_index$ or $sec\_unsorted\_array$.
To ensure the efficiency of the deletion, we do not convert the $sec\_learned\_index$ to $sec\_unsorted\_array$ even if the degree of a vertex is less than $T$.

\section{Experiments}
\label{sec:exp}
\subsection{Experimental Setup}
\noindent{{\bf Environment.}}
Our system is implemented in C++, and all experiments are conducted on a Linux with 2.1GHz Intel Xeon Silver 4110 CPU and 256G RAM. 
We set $T=60$ by default. 

\begin{table}
\begin{center}
\newcommand{\tabincell}[2]{\begin{tabular}{@{}#1@{}}#2\end{tabular}}
\caption{Statistics of graphs.}
\vspace{-0.3cm}
\resizebox{0.9\linewidth}{!}{
\begin{tabular}{c|cccc}
      \hline
      Graph & {\tabincell{c}{\# Vertices}} & {\tabincell{c}{\# Edges}} & {\tabincell{c}{Max. degree}} & {\tabincell{c}{Avg. degree}} \\
      \hline
      G500-24 & 8,870,942 & 260,379,521 & 406,416 & 58 \\
      G500-26 & 32,804,978 & 1,051,922,853 & 1,003,338 & 54 \\
      Orkut & 3,072,441 & 117,185,083 & 33,007 & 43 \\
      LiveJ & 3,997,962 & 34,681,189 & 14,815 & 17 \\
      \hline
    \end{tabular}}
\label{tab:dataset}
\end{center}
\vspace{-0.5cm}
\end{table}

\noindent\textbf{Competitors.}
We evaluate on six systems: \textbf{\emph{Teseo}\cite{Teseo}}, \textbf{{\emph{Sortledton~\cite{Sortledton}}}}, \textbf{\emph{LiveGraph~\cite{zhu2020livegraph}}}, \textbf{\emph{Aspen~\cite{dhulipala2019low}}}, \textbf{\emph{LSGraph~\cite{qi2024lsgraph}}}, and \textbf{\emph{LGstore}}.

\noindent\textbf{Datasets.} 
We evaluate on two synthetic datasets Graph500-24 (G500-24), Graph500-26 (G500-26) from LDBC Graphalytics Benchmark \cite{graphalytics-benchmark}, and two real-world datasets Orkut, LiveJournal (LiveJ). Table \ref{tab:dataset} shows the statistics of these datasets. 

\noindent\textbf{Transactional workloads.} We evaluate throughput for three workloads: write only (A), 50\% write and read (B) and read only (C).

\noindent\textbf{Graph analytics algorithms.} We evaluate five graph analytics algorithms from LDBC Graphalytics Benchmark \cite{graphalytics-benchmark}.

\begin{figure*}[t]
\centering
\includegraphics[width=0.7\linewidth]{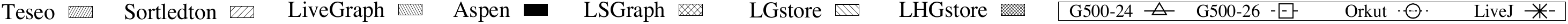}  
\vspace{-0.1cm}
\\

\subfigure[Workload A\label{subfig:write_basline}]{
    \includegraphics[width=0.148\textwidth]{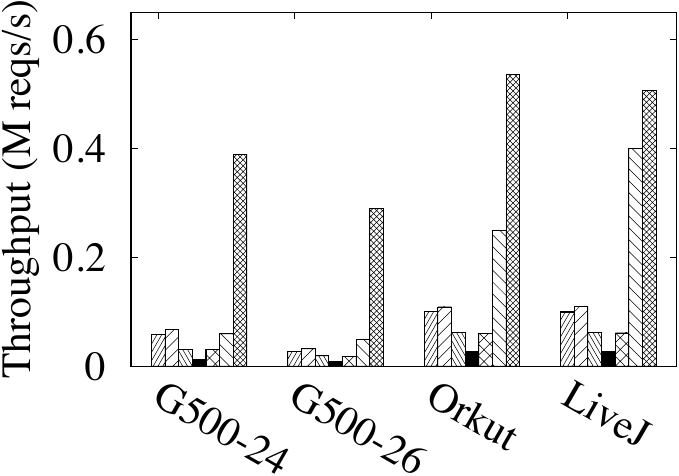}
}\hfill
\subfigure[Workload A\label{subfig:A_T_T}]{
    \includegraphics[width=0.16\textwidth]{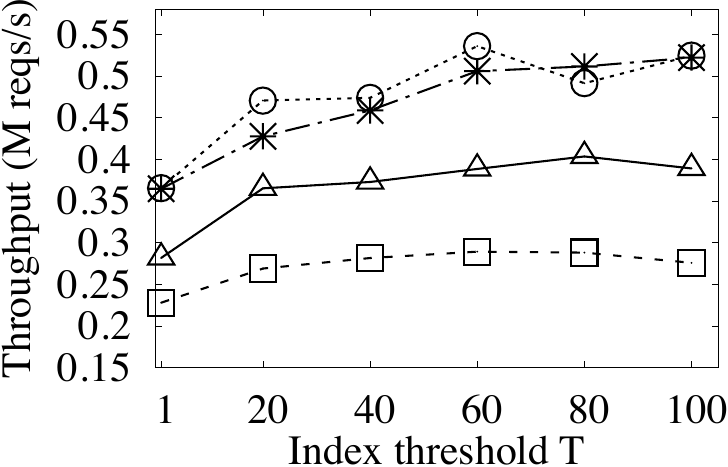}
}\hfill
\subfigure[Workload B\label{subfig:50write_basline}]{
    \includegraphics[width=0.148\textwidth]{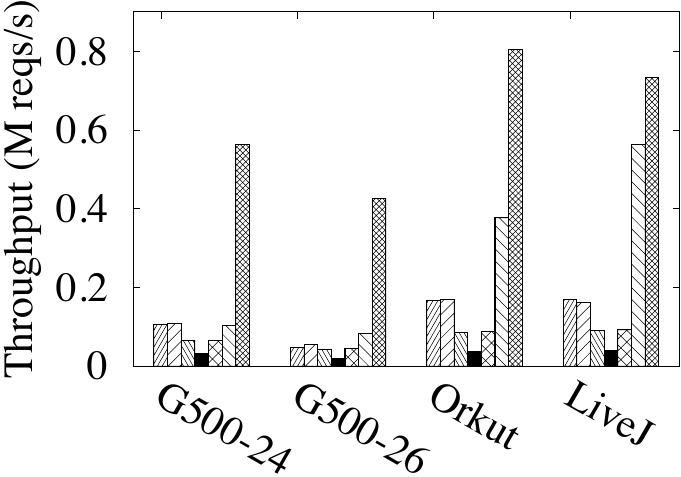}
}\hfill
\subfigure[Workload B\label{subfig:B_T_T}]{
    \includegraphics[width=0.16\textwidth]{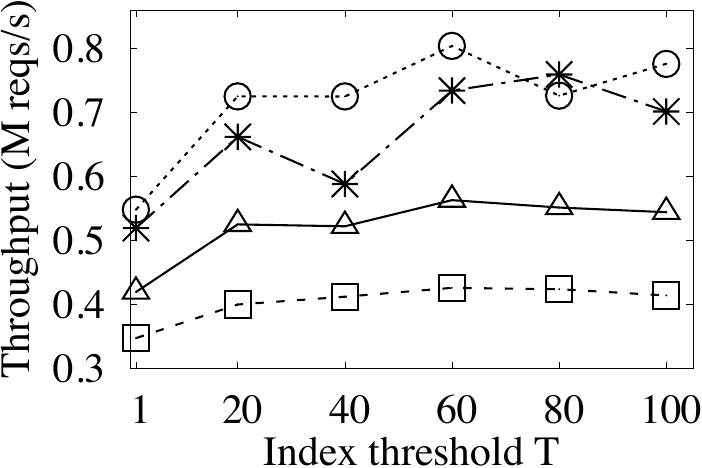}
}\hfill
\subfigure[Workload C\label{subfig:read_basline}]{
    \includegraphics[width=0.148\textwidth]{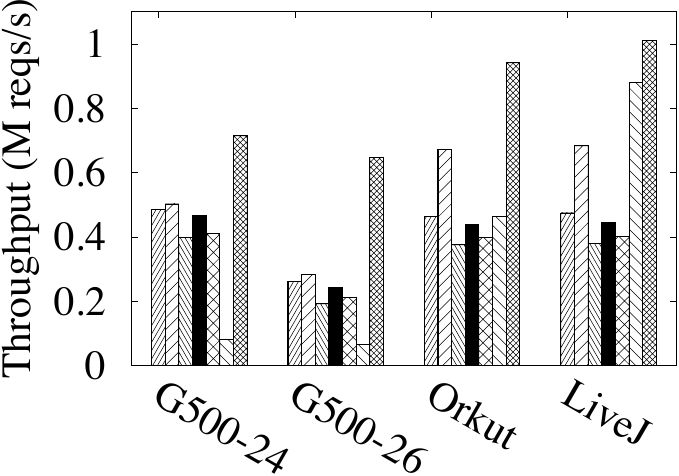}
}\hfill
\subfigure[Workload C\label{subfig:C_T_T}]{
    \includegraphics[width=0.16\textwidth]{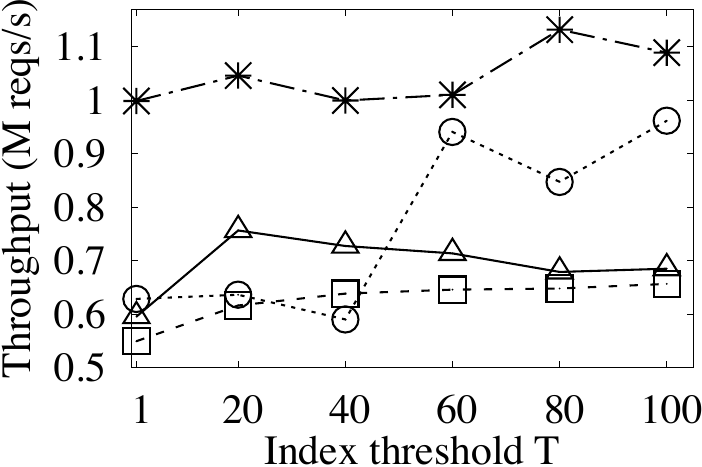}
}

\vspace{-0.4cm}
\caption{Comparison of throughput under Workload A, B, and C, and the effect of varying $T$.}
\label{fig:exp_epoch}
\vspace{-0.2cm}
\end{figure*}

\begin{table*}[t]
    \centering
    \caption{Comparison of running time on five graph algorithms across four datasets.}
    \vspace{-0.3cm}
    \label{tab:algo time}
    \resizebox{\linewidth}{!}{
    \begin{tabular}{c|cccc|cccc|cccc|cccc|cccc}
        \hline
        \multirow{2}{*}{\textbf{System}} 
        & \multicolumn{4}{c|}{\textbf{BFS}}
        & \multicolumn{4}{c|}{\textbf{PageRank}}
        & \multicolumn{4}{c|}{\textbf{LCC}}
        & \multicolumn{4}{c|}{\textbf{WCC}}
        & \multicolumn{4}{c}{\textbf{SSSP}} \\
        \cline{2-21}
        & LiveJ & Orkut & G500-24 & G500-26
        & LiveJ & Orkut & G500-24 & G500-26
        & LiveJ & Orkut & G500-24 & G500-26
        & LiveJ & Orkut & G500-24 & G500-26
        & LiveJ & Orkut & G500-24 & G500-26 \\
        \hline

        Teseo 
        & 48.7s & 286.1s & 586.4s & 3091.2s
        & 294.9s & 1909.6s & 3904.8s & 19297.9s
        & 0.57s & 11.4s & 212.3s & 1216.3s
        & 55.9s & 287.9s & 614.1s & 3133.6s
        & 204.2s & 1413.1s & 3062.3s & 15550.2s \\

        Sortledton 
        & 25.1s & 341.6s & 562.3s & 2984.9s
        & 498.2s & 2827.3s & 4792.6s & 28443.5s
        & 0.65s & 11.4s & 215.0s & 1250.0s
        & 65.2s & 211.7s & 673.2s & 3410.5s
        & 152.4s & 1139.7s & 2591.2s & 12249.8s \\

        LiveGraph 
        & 69.2s & 823.4s & 987.3s & 5630.7s
        & 997.4s & 4725.8s & 6015.8s & 41781.3s
        & 0.82s & 16.5s & 254.7s & 1565.3s
        & 110.5s & 544.1s & 1123.2s & 5922.7s
        & 298.7s & 2397.4s & 3577.9s & 19441.2s \\

        Aspen 
        & 24.5s & 278.2s & 548.7s & 2967.6s
        & 517.3s & 2180.6s & 4438.2s & 30028.8s
        & 0.70s & 13.1s & 230.3s & 1309.2s
        & 81.3s & 197.3s & 635.4s & 3045.5s
        & 233.1s & 1299.3s & 2824.5s & 13912.8s \\

        LSGraph 
        & 22.8s & 273.2s & 540.5s & 2915.5s
        & 388.2s & 2279.1s & 4297.7s & 24883.1s
        & 0.69s & 16.5s & 254.7s & 1656.3s
        & 56.8s & 272.8s & 601.2s & 3008.2s
        & 97.8s & 951.4s & 2498.2s & 12773.4s \\

        LGstore 
        & 12.1s & 29.4s & 105.0s & 622.5s
        & 112.6s & 297.9s & 1212.1s & 7834.1s
        & 0.67s & 17.3s & 1661.9s & 15977.1s
        & 13.9s & 28.8s & 119.4s & 681.1s
        & 47.2s & 130.3s & 459.6s & 3302.4s \\

        LHGstore 
        & \textbf{8.8s} & \textbf{24.5s} & \textbf{106.1s} & \textbf{619.2s}
        & \textbf{73.1s} & \textbf{204.8s} & \textbf{805.4s} & \textbf{5164.5s}
        & \textbf{0.27s} & \textbf{4.9s} & \textbf{109.9s} & \textbf{521.1s}
        & \textbf{11.1s} & \textbf{26.7s} & \textbf{114.3s} & \textbf{577.7s}
        & \textbf{39.6s} & \textbf{125.9s} & \textbf{434.9s} & \textbf{2913.5s} \\
        \hline
    \end{tabular}}
    \vspace{-0.2cm}
\end{table*}

\begin{figure*}[!t]
\centering
\includegraphics[width=0.3\linewidth]{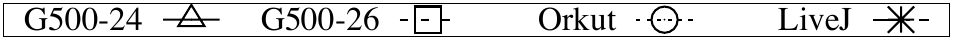}
\vspace{-0.1cm}

\subfigure[BFS\label{subfig:BFS_T}]{
    \includegraphics[width=0.175\linewidth]{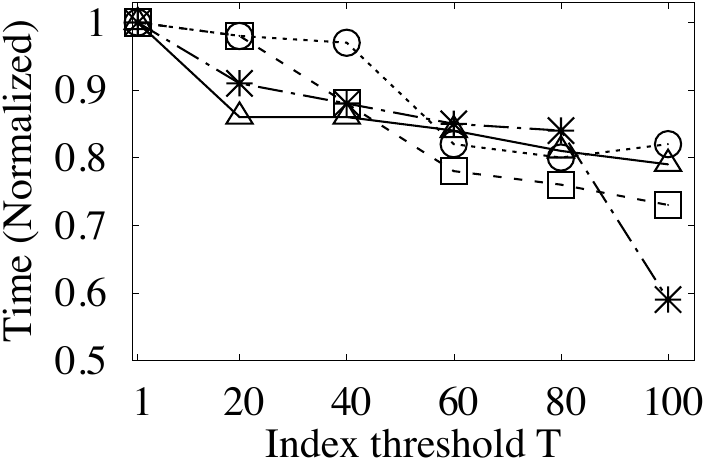}
}\hfill
\subfigure[PageRank\label{subfig:PR_T}]{
    \includegraphics[width=0.175\linewidth]{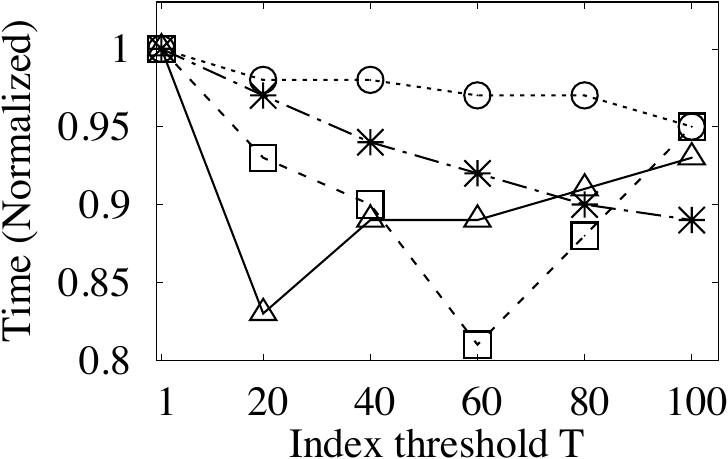}
}\hfill
\subfigure[LCC\label{subfig:LCC_T}]{
    \includegraphics[width=0.175\linewidth]{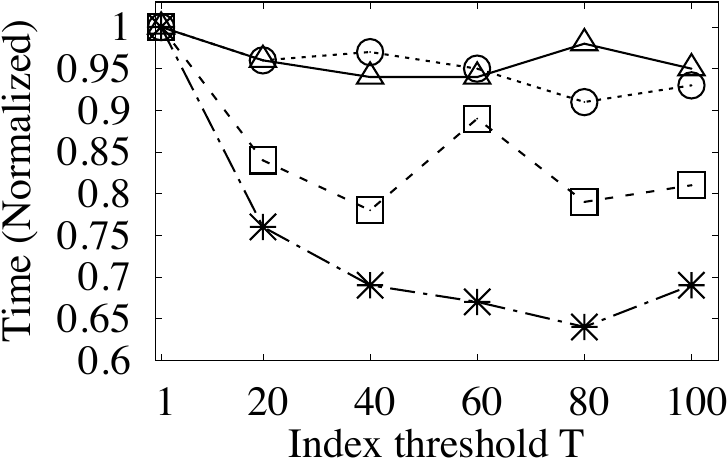}
}\hfill
\subfigure[WCC\label{subfig:WCC_T}]{
    \includegraphics[width=0.175\linewidth]{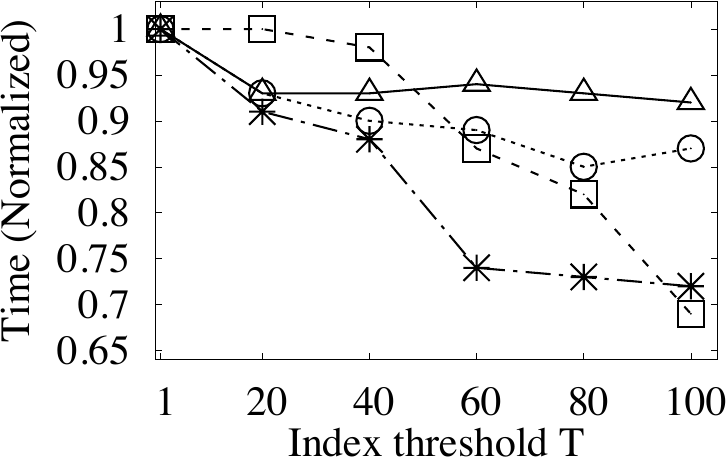}
}\hfill
\subfigure[SSSP\label{subfig:SSSP_T}]{
    \includegraphics[width=0.175\linewidth]{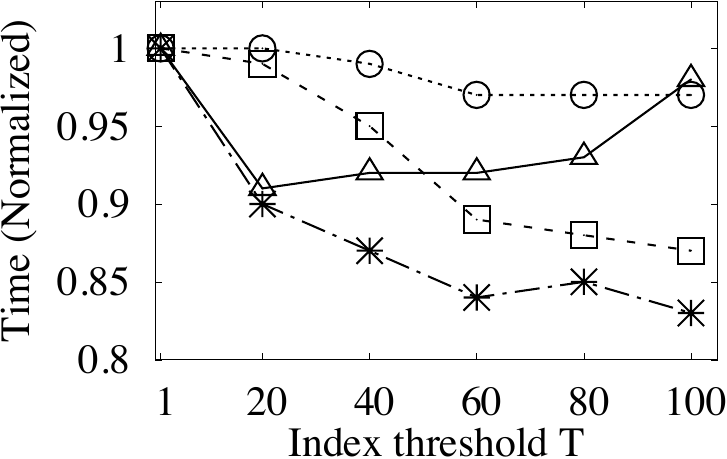}
}

\vspace{-0.4cm}
\caption{Running time (normalized to the runtime when $T=1$) of five algorithms under different $T$.}
\vspace{-0.4cm}
\label{fig:exp_algorithrm_T}
\end{figure*}

\subsection{Update Throughput}
Fig.~\ref{subfig:write_basline}, \ref{subfig:50write_basline} and \ref{subfig:read_basline} report the throughput under the three workload. Across all datasets, LHGstore achieves the highest throughput in all three settings. Under Workload A, LHGstore improves over LGstore by 6.6$\times$, 5.9$\times$, 2.2$\times$, and 2.2$\times$ on G500-24, G500-26, Orkut, and LiveJ, and exceeds the slowest baselines by up to 12.1–28.2$\times$. Under Workload B, LHGstore achieves 1.3–5.4$\times$ higher throughput than LGstore and as much as 9.3–20.1$\times$ higher than systems that rely on skip lists, logs, or versioned structures. Even under Workload C, where locality plays a more important role, LHGstore still outperforms LGstore by 1.1–9.5$\times$ and the slowest baselines by 5.9–17.4$\times$, mainly because the learned edge index avoids linear scans on high-degree vertices. 
These results show that LHGstore minimizes expensive pointer chasing on high-degree vertices while preserving contiguous scans on low-degree vertices, which translates directly into higher throughput on real hardware.

\subsection{Analytics Performance}

We evaluate analytics performance on five graph algorithms, as shown in Table \ref{tab:algo time}, grouped by their dominant access patterns.

\noindent\textbf{Random vertices with sequential neighbor access.}
BFS, WCC, and SSSP fall into this category, where each iteration touches one vertex and scans its adjacency list. LHGstore achieves the best or near-best performance, achieving 5.5–11.7$\times$ speedups over Teseo and 2.6–13.9$\times$ over Sortledton and LSGraph. LGstore performs well due to its contiguous layout, occasionally approaching LHGstore in purely traversal-heavy workloads. However, unlike LGstore, LHGstore sustains this performance without compromising update efficiency. Teseo is slowed by expensive ART lookups, while LiveGraph and Aspen suffer from pointer-heavy that degrades locality.

\noindent\textbf{Sequential vertices and neighbor access.}
PageRank scans all vertices and adjacency lists sequentially. LHGstore outperforms all baselines because it records vertex degrees directly in edge blocks, enabling efficient sequential traversal. LSGraph mitigates skew but lacks prediction, and Aspen and LiveGraph incur higher costs due to tree- or log-based traversal.

\noindent\textbf{Random neighbor access.}
LCC involves random neighbor checks. LGstore performs poorly because it relies on linear scans. LHGstore’s learned index avoids such scans and achieves 2.4–30.6$\times$ speedups over LGstore while outperforming all other baselines. LiveGraph and Aspen are particularly slow because random neighbor checks repeatedly traverse logs or trees.

\subsection{Effect of Index Threshold}
To evaluate the benefit of assigning different storage structures to vertices of different degrees, we examine how the index threshold $T$ affects update throughput and analytics performance. As shown in Fig.~\ref{subfig:A_T_T}, \ref{subfig:B_T_T}, and \ref{subfig:C_T_T}, all configurations with $T>1$ substantially outperform $T$=1, confirming that separating low-degree and high-degree vertices is crucial for skewed graphs. Across most datasets, throughput peaks around $T=60$. Occasional fluctuations arise because real-world degree distributions are discrete; when $T$ crosses a dense degree region, many vertices change layouts simultaneously, causing abrupt performance shifts.

For graph analytics, Fig.~\ref{subfig:BFS_T}–\ref{subfig:SSSP_T} normalize running times to the $T=1$ baseline. In general, larger $T$ improves performance since low-degree vertices benefit from contiguous arrays during traversal. However, algorithms such as LCC and PageRank do not always monotonically improve because their runtimes depend on both sequential traversal (favoring unsorted arrays) and random adjacency checks (favoring learned index). Overall, choosing an appropriate $T$ enables LHGstore to support high-throughput graph updates and high-performance graph analysis.

\begin{figure}
\vspace{-0.1cm}
 \centering
 \begin{tabular}[t]{l}
    \hspace{-0.3cm}
  \subfigure[]{
   \label{subfig:mem}
   \raisebox{0.03\height}{\includegraphics[width=0.4\linewidth]{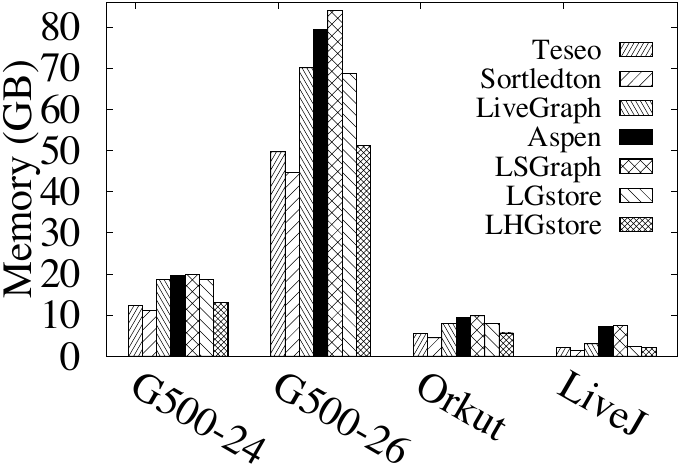}}
  }
      \hspace{0.3cm}
  \subfigure[]{
   \label{subfig:mem_T}
   \raisebox{0.03\height}{\includegraphics[width=0.4\linewidth]{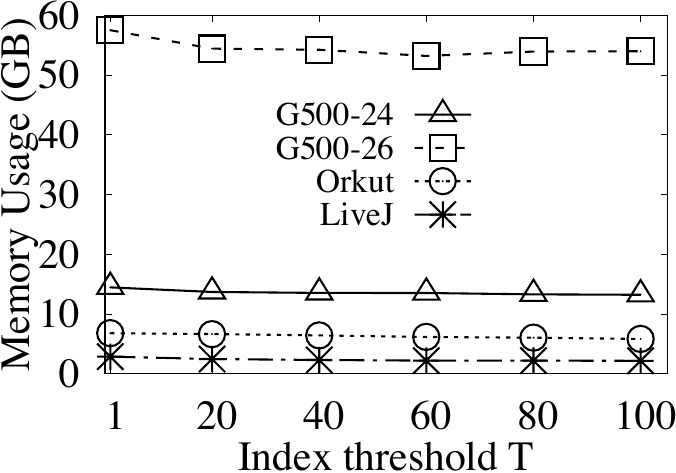}}
  }
 \end{tabular}
 \vspace{-0.5cm}
 \caption{(a) Comparison of the memory usage and (b) the memory usage of LHGstore under different $T$.}
 \label{fig:mem usage}
 \vspace{-0.5cm}
\end{figure}

\subsection{Memory Usage}
Fig.~\ref{subfig:mem} compares the memory usage of all systems. Overall, LHGstore achieves competitive efficiency, close to Teseo and lower than most other baselines. The extra metadata in LHGstore’s edge blocks (e.g., pointers for the second-level index) introduces moderate overhead on large graphs such as G500-26, but it is still smaller than LiveGraph’s logs, Aspen’s versioned structures, or Sortledton’s skip lists. 
LGstore consumes more memory because it duplicates source vertices for each edge.
In addition, the memory usage for different $T$ is shown in Fig. \ref{subfig:mem_T}. As $T$ increases, the memory usage of LHGstore decreases slightly because more vertices use unsorted arrays, which are more space-efficient than learned-index blocks.

\section{Conclusion}  
\label{sec:conclusion}
We presented LHGstore, a degree-aware learned hierarchical graph storage framework for in-memory dynamic graphs. By decoupling vertex and edge access and assigning adjacency layouts based on vertex degree, LHGstore reconciles the trade-off between update efficiency and traversal locality. Experiments show that LHGstore achieves substantial improvements in both update throughput and analytics performance over SOTA in-memory graph systems. 

\bibliographystyle{ACM-Reference-Format}
\bibliography{sample}

\end{document}